\documentclass[twocolumn,aps,pra,showpacs,superscriptaddress]{revtex4}
\usepackage[usenames,dvipsnames]{color}

\usepackage{graphicx}
\usepackage{amsmath,amssymb}
\usepackage{color}
\usepackage{graphics}
\usepackage{epsfig}
\usepackage{hyperref}

\newcommand{\be}{\begin{equation}}
\newcommand{\ee}{\end{equation}}

\newcommand{\bea}{\begin{eqnarray}}
\newcommand{\eea}{\end{eqnarray}}

\newcommand{\eins}{\openone}
\newcommand{\qed}{\ensuremath{\hfill \Box}}
\newcommand{\WW}{\ensuremath{\mathcal{W}}}

\newcommand{\NN}{\ensuremath{\mathcal{N}}}

\newcommand{\FF}{\ensuremath{\mathcal{F}}}

\newcommand{\ketbra}[1]{\ensuremath{| #1 \rangle \!\langle #1 |}}

\newcommand{\ket}[1]{\ensuremath{|#1\rangle}}

\newcommand{\bra}[1]{\ensuremath{\langle#1|}}

\newcommand{\braket}[2]{\ensuremath{\langle #1|#2\rangle}}

\newcommand{\kommentar}[1]{}

\renewcommand{\vr}{\ensuremath{\varrho}}

\newcommand{\forget}[1]{}

\begin{document}


\title{Evaluating the geometric measure of 
multiparticle entanglement}



\author{Lars Erik Buchholz}
\affiliation{Naturwissenschaftlich-Technische Fakult\"at,
Universit\"at Siegen,
Walter-Flex-Str.~3,
D-57068 Siegen}

\author{Tobias Moroder}
\affiliation{Naturwissenschaftlich-Technische Fakult\"at,
Universit\"at Siegen,
Walter-Flex-Str.~3,
D-57068 Siegen}

\author{Otfried G{\"u}hne}
\affiliation{Naturwissenschaftlich-Technische Fakult\"at,
Universit\"at Siegen,
Walter-Flex-Str.~3,
D-57068 Siegen}


\date{\today}


\begin{abstract}
We present an analytical approach to evaluate the geometric measure 
of multiparticle entanglement for mixed  quantum states. Our method 
allows the computation of this measure for a family of multiparticle 
states with a certain symmetry and delivers lower bounds on the measure
for general states. It works for an arbitrary number of particles,
for arbitrary classes of multiparticle entanglement, and can also be 
used to determine other entanglement measures.
\end{abstract}




\maketitle


\section{Introduction}
The characterization of quantum correlations is central 
for many applications such as quantum metrology or quantum 
simulation. Moreover, the quantification of quantum correlations
allows to study quantum phase transitions and is also relevant 
for quantum optics experiments, where coherent dynamics of many 
particles should be certified.  This characterization, however, 
is difficult as for multiparticle systems different forms of 
entanglement exist. To give a simple example, a quantum state 
$\ket{\psi}$ on three particles can be fully separable, 
$\ket{\psi_{fs}}=\ket{a}\otimes\ket{b}\otimes\ket{c}$, 
or entangled on two particles, but separable for the third 
one. An instance of such a biseparable state is 
$\ket{\psi_{bs}}=\ket{\phi_{ab}}\otimes\ket{\eta_c}$, which is
biseparable with respect to the $AB|C$-partition, but there 
are two other possible bipartitions. Finally, if a state is not 
fully separable or biseparable, it is called genuine multiparticle 
entangled. For more than three particles even more classes are possible.
Here, a pure state is called $k$-separable, if the $N$ particles can 
be split into $k$ unentangled groups. The case $k=N$ corresponds to the 
fully separable states, while $k=2$ denotes the biseparable states \cite{gtreview}. 

So far, this is only a classification for pure states. In order 
to study multiparticle entanglement in realistic situations, two
extensions are necessary: First, one needs to deal with mixed states, 
as they occur naturally in experiments. Second, one needs to {\it quantify}
entanglement in order to go beyond the simple entangled vs.~separable 
scheme \cite{measure-reviews}. A popular way to quantify multiparticle entanglement makes 
use of the geometric measure of entanglement \cite{wei}. For that, one 
defines for pure states via
\be
E_G^{(k)}(\ket{\psi}) = 1 - \max_{\ket{\phi} {\rm \; is \\ \;\; k-sep.}} 
|\braket{\phi}{\psi}|^2
\ee
the amount of entanglement as one minus the maximal overlap over pure
$k$-separable states. Clearly, this vanishes for $k$-separable pure
states and the measure is large for highly entangled states where the 
overlap is small. Note that the quantity $E_G^{(k)}$ distinguishes 
between the different classes of $k$-separability, so it is  indeed a 
whole family of entanglement measures suited for all the different 
entanglement classes \cite{illuminati}. The geometric measure found many applications, 
it can be used to characterize the distinguishability of quantum 
states by local means \cite{geo-apps} and to investigate quantum phase transitions 
in spin models \cite{geo-spin}, to mention only a few. Moreover, it is directly 
linked to other entanglement measures, which quantify the distance 
to the separable states \cite{streltsov}.

The quantity needs to be extended to mixed quantum states. For that, 
one uses the so-called convex roof construction. One defines
\be
E_G^{(k)}(\vr) = \min_{p_i, \ket{\psi_i}} \sum_i p_i E_G^{(k)}(\ket{\psi_i}),
\ee
where the minimization is taken over all possible decompositions 
$\vr = \sum_i p_i \ketbra{{\psi_i}}$ into pure states. This is a 
typical method to extend a pure state entanglement measure to the mixed
state case, but obviously this minimization is difficult to perform. So 
far, only in some special cases the evaluation of the geometric measure 
for mixed states is possible \cite{wei, gbb, indiaposter}. Analytical results on the evaluation 
of the convex roof for other measures have to date mainly 
be obtained for the bipartite measure of the concurrence and measures 
with a related mathematical structure \cite{wootters}. Further results are 
typically restricted  to special families of states or special 
entanglement  classes \cite{eltschka, furtherexact} or require a numerical optimization 
\cite{geza-neu}.

In this paper, we present an analytical way to evaluate the geometric
measure of entanglement for mixed states of $N$ qubits. Our approach 
gives the exact value for a class of states, 
for other states the approach results in a lower bound on the measure. 
The method works for an arbitrary number of qubits and arbitrary 
kinds of $k$-separability. Since it has been shown that computing 
the geometric measure is equivalent to determining other distance
related entanglement measures \cite{streltsov}, our results constitute 
one of the most detailed characterizations of multiparticle entanglement
so far.

\begin{figure}[t!]
\begin{center}
\includegraphics[width=0.9\columnwidth]{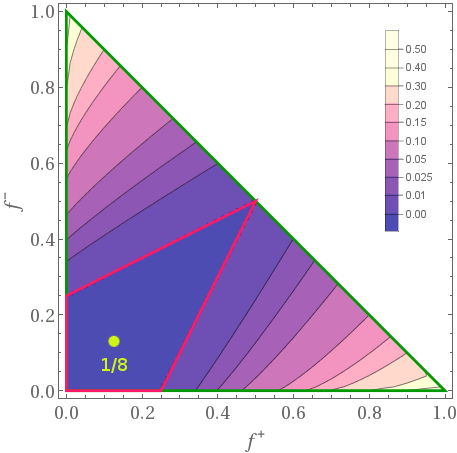}
\end{center}
\caption{The triangle of the GHZ-symmetric states for three qubits.
The states are parameterized by the two fidelities $f^+$ and $f^-$
and include besides the GHZ states also the maximally mixed state 
$\eins/8$. The values of the geometric measure of entanglement $E_G^{(3)}$
are shown as contours, and the red deltoid denotes the parameter
region where the states are fully separable, so the geometric measure
vanishes.}
\label{eg-bild-4}
\end{figure}

\section{GHZ-symmetric states}
We start by defining a two-parameter family of states that we use for 
our study.
Consider three qubits and the states
\begin{align}
\vr^{\rm ghz}&(f^+, f^-) = 
f^+ \ketbra{GHZ^+} 
\nonumber
\\
&+ f^- \ketbra{GHZ^-} + (1-f^+ - f^-) \frac{\Pi}{6},
\label{ghzsymm-3}
\end{align}
where $\ket{GHZ^\pm} = (\ket{000}\pm \ket{111})/\sqrt{2}$ are 
two Greenberger-Horne-Zeilinger (GHZ) states,
$\Pi=\eins-\ketbra{GHZ^+}-\ketbra{GHZ^-}$ is the projector onto 
the space orthogonal to the GHZ states, and $f^\pm \in [0,1]$ with 
$f^+ + f^- \leq 1$. This family of states 
forms a triangle in the state space (see also Fig.~\ref{eg-bild-4})
and includes besides the GHZ states also the maximally mixed
state $\eins/8=\vr^{\rm ghz}(1/8,1/8).$ We note that states of 
this form have been studied before and recently, Eltschka and
Siewert succeeded by computing the three-tangle as an entanglement
measure for them \cite{eltschka}. In the following, we will call this family
of states GHZ-symmetric states.

Before discussing our main results, we note some symmetries of the states. 
We have
\be
\vr^{\rm ghz}= U \vr^{\rm ghz} U^\dagger,
\ee
where either (i) $U$ induces a permutation of the qubits or 
(ii) $U = \sigma_x \otimes\sigma_x \otimes\sigma_x$ is a 
spin-flip operation on all three qubits  or 
(iii) $U$ describes correlated qubit rotations of the 
form $U(\alpha, \beta, \gamma) = e^{i\alpha \sigma_z} 
\otimes e^{i\beta \sigma_z} \otimes e^{i\gamma \sigma_z}$
with $\alpha +\beta + \gamma = 0.$
In fact, one can easily see that the states $\vr^{\rm ghz}$ are 
the only states invariant under these operations \cite{eltschka}.

{From} this it follows that the states $\vr^{\rm ghz}$ have, 
for any convex entanglement measure obeying the additional 
constraint of being permutation-invariant \cite{horosym},
the minimal entanglement among all states with the same GHZ 
fidelities $f^+$ and $f^-$. This can be seen as follows: 
The transformations (i), (ii) and (iii) from above do not 
change the amount of entanglement. So, if an arbitrary state $\vr$ is 
symmetrized with respect to a given unitary, 
$\vr\mapsto (\vr+ U \vr U^\dagger)/2$, the entanglement
measure will not increase due to its convexity. After 
symmetrization with respect to all three transformations
the state $\vr$ is made GHZ-symmetric. The fidelities $f^+$ 
and $f^-$ are invariant under the symmetrization
since the GHZ states $\ket{GHZ^\pm}$ are invariant, but the 
entanglement is decreasing, which proves the claim.

This minimality is not only convenient for computing
the geometric measure, it is also useful for experiments: An 
experimenter who has not full information about the quantum state 
may only measure the fidelities $f^+$ and $f^-$ and compute
the entanglement of the GHZ-symmetric states with our formulas 
below. The resulting value is automatically a lower bound on the
different forms of multiparticle entanglement present in the 
experiment. If more information on the state is available, 
one may even optimize $f^+$ and $f^-$ under local
transformations \cite{eltschka2}.

\section{Results for three qubits}
We consider the geometric measure for full separability 
$E_G^{(3)}$, where the overlap with fully separable states 
is taken. The corresponding measure for genuine multiparticle
entanglement will be considered later. We can directly state:

{\bf Observation 1.} Let $\vr^{\rm ghz}$ be a GHZ-symmetric 
state of three qubits. Then, the geometric measure is given by
\be
E_G^{(3)}= 
\!\!
\max_{\mu \in [0,1]}
\frac{1}{2}
\big[
1+\mu (2 f^+ - 1)
-\sqrt{\alpha}+\frac{f^- \mu (\mu+\sqrt{\alpha})}{\mu-1}
\big],
\label{finalopti-o1}
\ee
where we used the abbreviation $\alpha=1-\mu+\mu^2.$ This formula
holds for $f^+ \geq f^-$, for the other case one has to exchange the
values of $f^+$ and $f^-$.

The idea of the proof is the following: The entanglement $E_G^{(3)}$ 
is the minimal entanglement compatible with the fidelities $f^+$ and 
$f^-$. As the measure is defined via the convex roof, the minimal 
entanglement is convex, too. This convex function can be expressed 
with the help of the Legendre transformation \cite{legendretrafo}. So one has to 
compute the Legendre transform of the geometric measure, and from 
that one can compute the geometric measure itself. The detailed proof, 
however, involves some nontrivial optimizations and the use of criteria 
for full separability of quantum states and is given in the Appendix.

To start the discussion, one may wonder 
why the computation of the geometric measure still contains a 
remaining optimization. The main reason is notational simplicity:
The remaining optimization can analytically be evaluated for any 
given $f^+, f^-$ by setting the derivative with respect to $\mu$
equal zero. In fact, even a closed formula for arbitrary  $f^+, f^-$  
is possible \cite{compi}, the final result, however, is quite lengthy 
and does not lead to new insights. In Fig.~\ref{eg-bild-4} we show the 
value of the geometric measure according to Observation 1. Note that for 
$f^+ \leq (2f^- +1)/4$ and $f^-\leq (2f^+ +1)/4$ the state is PPT and 
hence fully separable. This was was known before \cite{duer01, kay}, extensions of 
this statement will be discussed below.

For important subfamilies of GHZ-symmetric states, closed formulas can
directly be written down. First, one may be interested in the hypotenuse 
of the triangle, where $f^+ + f^- =1.$ For this case and $f^+\geq f^-$ 
our method gives
\be
E_G^{(3)}=\frac{1}{2}(1-\sqrt{1-(2f^+-1)^2}),
\label{upper-edge}
\ee
in agreement with previous results \cite{gbb}.
Second, for the lower cathetus (where $f^-=0$) one finds that
\bea
E_G^{(3)} &=& \frac{1}{4}\big[1+2f^+-2\sqrt{3}\sqrt{f^+(1-f^+)}\big]
\nonumber
\\
&& \mbox{ for } f^+ \in [1/4,3/4],
\nonumber
\\
E_G^{(3)} &=& f^+-1/2 \;\;\; \mbox{ for } f^+ \in [3/4, 1].
\label{lowerborder}
\eea
Given the values of $E_G^{(3)}$ it is interesting to ask about the 
structure and form of the optimal decomposition in the convex roof.
This does not only demonstrate the complete understanding of the 
problem, it can also be used to determine the closest separable 
state with respect to the Uhlmann fidelity \cite{streltsov}. We find 
that the  optimal decomposition for the lower cathetus for 
$f^+ \in [1/4,3/4]$ consists of 28 states. One of them is the 
state
\begin{align}
\ket{\psi_1} &= \alpha (\ket{000}+\ket{111})
\nonumber
\\
&+ \beta (\ket{001}+\ket{010}+\ket{011}+\ket{100}+\ket{101}+\ket{110})
\end{align}
with $\alpha = \sqrt{f^+/2}$ and $\beta=\sqrt{(1-2\alpha^2)/6}
=\sqrt{(1-f^+)/6}$ and the other vectors are obtained by applying 
transformations of the type $e^{i \phi_1 \sigma_z} \otimes e^{i\phi_2 \sigma_z} 
\otimes e^{i\phi_3 \sigma_z}$ to $\ket{\psi_1}.$ For $f^+ \in [3/4,1]$
the optimal decomposition is of the type $\vr(f^+)=(1-p) \vr^{\rm ghz}(3/4,0) + 
p \ketbra{GHZ^+}$, explaining the linear behavior of $E_G^{(3)}$.
The details are given in the Appendix.

\section{The case of $N$ qubits and $k$-separability}
We can easily extend the family of GHZ-symmetric states
to an arbitrary number of qubits by replacing $\ket{GHZ^+}$ and $\ket{GHZ^-}$
with the corresponding $N$-qubit GHZ states and considering
\bea
\vr^{\rm ghz} &=& 
f^+ \ketbra{GHZ^+_N} + f^- \ketbra{GHZ^-_N}
\nonumber
\\
&&+ (1-f^+ - f^-) \frac{\Pi_N}{2^N-2}
\label{ghzsymm-n}
\eea
where $\Pi_N$ is now a projector onto a the $2^N-2$-dimensional space 
orthogonal to the GHZ states. This family of states obeys analogous 
symmetries as the three-qubit states. The reader may 
notice at this point that the $N$-qubit GHZ-symmetric states are not
uniquely determined by the symmetries. This means that it is not clear 
from the beginning that a lower bound based on the fidelities $f^+$ and
$f^-$ gives the exact value of the entanglement. We will see later that 
this is nevertheless the case, for the moment we just state:

{\bf Observation 2.}
Let $\vr^{\rm ghz}$ be an $N$-qubit GHZ-symmetric quantum state and 
$E_G^{(k)}$ the geometric measure with respect to 
$k$-separability with $3 \leq k \leq N$. Then, this measure 
is given by:
\begin{align}
&E_G^{(k)}(f^+, f^-) 
\label{finaloptimulti}
\\
&=\max_{\mu \in [0,\mu_{\rm max}]}
\frac{1}{2}
\big[
1+\mu (2 f^+ - 1)
-\sqrt{\gamma}+\frac{f^- \mu (\mu+\sqrt{\alpha})}{\mu-1}
\big],
\nonumber
\end{align}
where we have used
$\mu_{\rm max}=2^{k-3}/(2^{k-2}-1)$
and
$\gamma=(\mu-1)^2+2^{3-k}\mu$
and 
$\alpha=1-\mu+\mu^2.$ 
The proof is given in the Appendix.

\begin{figure}[t]
\begin{center}
\includegraphics[width=0.9\columnwidth]{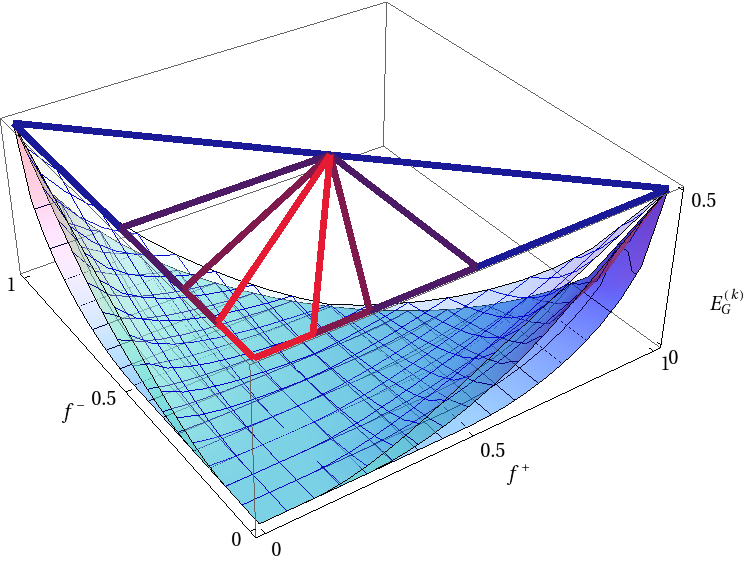}
\end{center}
\caption{The geometric measure of entanglement for four qubits. 
From top to bottom, the different values for full separability 
($k=4$), three-separability ($k=3$) and biseparability ($k=2$) 
are shown. The square and the deltoids on the top denote 
the regions, where the measures vanish.}
\label{eg-bild-3}
\end{figure}

Note that this formula does not depend on the number of qubits,
but on the parameter $k$ characterizing the separability. Also, 
for the case $k=3$ we have $\gamma=\alpha$, and the expression 
reduces to Eq.~(\ref{finalopti-o1}).
Finally, it remains to consider the case $k=2$, i.e. the geometric 
measure for genuine multiparticle entanglement. In this case, we 
obtain a very simple solution:

{\bf Observation 3.} The geometric measure of genuine multiparticle
entanglement is for arbitrary GHZ-symmetric states given by
\be
E_G^{(2)}(f^+, f^-) = \frac{1}{2}-\sqrt{f(f-1)},
\ee
where $f = \max\{f^+,f^-\}$ is the larger of the two fidelities.
The formula is valid if $f\geq 1/2$, otherwise the state is biseparable.
The proof is given in the Appendix.

Equipped with these results, we can visualize the geometric measure for 
the different classes, an example is provided in Fig.~\ref{eg-bild-3}.
We add that from our proof it follows that an arbitrary $k$-separable 
multi-qubit state has to fulfill the conditions
\bea
f^+ \leq \frac{1}{2^{k-1}}\big[1+(2^{k-1}-2){f^-}\big],
\nonumber
\\
f^- \leq \frac{1}{2^{k-1}}\big[1+(2^{k-1}-2){f^+}\big],
\label{alphacondition2-maintext}
\eea
see also Eq.~(\ref{alphacondition2}) in the Appendix. Applied to the
GHZ-symmetric states, these equations describe deltoid curves where the 
states are $k$-separable and where the corresponding measures vanish. 
These are also shown in Fig.~\ref{eg-bild-3}.

\section{Connections to other entanglement measures}
An interesting feature of the geometric measure is the fact that 
it is intimately connected to other measures \cite{streltsov, martin}. 
This allows to determine other measures from our results on the 
geometric measure. More importantly, however, this connection enables
us to prove properties of the geometric measure.

Let us first recall the central result of Ref.~\cite{streltsov}. The Uhlmann 
fidelity between two mixed states is given by 
\be
F(\vr, \sigma) = [Tr(\sqrt{\sqrt{\vr} \sigma \sqrt{\vr}})]^2 
= [Tr(|\sqrt{\vr}\sqrt{\sigma}|)]^2.
\ee
This can been used to define several types of distance-based entanglement
measures, such as the Bures measure of entanglement
$
E_B(\vr)= \min_{\sigma \in S} [2-2 \sqrt{F(\vr,\sigma)}],
$
where $S$ denotes the set of separable states, or the Groverian measure 
of entanglement, $E_{GV}(\vr)= \min_{\sigma \in S} \sqrt{1- F(\vr,\sigma)}.$
In Ref.~\cite{streltsov} it was shown that the geometric measure for mixed 
states obeys
\be
E_G (\vr) = 1-\max_{\sigma \in S}F(\vr,\sigma).
\label{alex}
\ee
This means that calculating the convex roof of the geometric measure
is equivalent to finding the closest separable state, which is a remarkable 
connection between two seemingly different optimization problems. It also 
shows that our results on the geometric measure can directly be used to 
determine the distance based measures mentioned above.

In order to use this result for a better understanding of the geometric 
measure, we note some properties of the Uhlmann fidelity \cite{karol}. 
First, it is unitarily invariant, $F(U \vr U^\dagger, U \sigma U^\dagger) 
= F(\vr, \sigma)$. Then, for a given $\vr$ the quantity $F(\vr, \sigma)$ is 
concave in $\sigma.$ Finally, it is monotonous under quantum operations, 
$F(\Lambda[\vr], \Lambda[\sigma]) \geq F(\vr, \sigma)$ for any completely 
positive map $\Lambda$. {From} this property and Eq.~(\ref{alex}) it 
follows that the geometric measure for a state $\vr$ decreases under a
positive map $\Lambda$, if the nearest separable state $\sigma$ is mapped
onto a separable states, i.e. $E_G[\Lambda(\vr)] \leq E_G(\vr)$ if
$\Lambda(\sigma)$ is separable. It also implies that the closest separable
state to a GHZ-symmetric state is also GHZ-symmetric.  Now we are 
able to show two results on the geometric measure:

First, we can see why the multi-qubit states in Eq.~(\ref{ghzsymm-n}) 
minimize the geometric measure among all states with the fidelities 
$f^+$ and $f^-$, although they are not uniquely determined by the 
GHZ symmetry. Applying the GHZ symmetrization operations to an arbitrary
quantum state can only decrease the value of $E_G$, so the state with the 
minimal $E_G$ and given fidelities will definitely be GHZ-symmetric. This 
means that the state looks already similar to the state in 
Eq.~(\ref{ghzsymm-n}), but the diagonal elements 
$\vr_{2,2}, \dots, \vr_{2^N-1, 2^N-1}$ 
do not have to be identical. More precisely, for three qubits 
they have to 
be identical as in Eq.~(\ref{ghzsymm-3}), but for four qubits there are 
two possibilities, 
depending on whether the number of excitations $\ketbra{1}$ in  $\vr_{ii}$ is 
even or odd. In any case, for this type of states the separability properties 
for fixed bipartitions are determined by the criterion of the positivity of the
partial transpose \cite{duer01} and the separability deltoids as in 
Figs.~\ref{eg-bild-4},\ref{eg-bild-3} and Eq.~(\ref{alphacondition2-maintext}),
are still appropriate, see also the proof of Observation 2. We can consider 
a global permutation of the computational basis, which permutes  
$\vr_{2,2}, ..., \vr_{2^N-1, 2^N-1}$ 
but leaves $\vr_{1,1}$ and $\vr_{2^N, 2^N}$ invariant. This is a unitary
positive map, which for GHZ-symmetric states also maps separable states 
to separable states. It follows that the closest separable state is mapped
to a separable state and so the geometric measure decreases, 
moreover, by applying a random permutation, the diagonal elements 
$\vr_{2,2}, ..., \vr_{2^N-1, 2^N-1}$ will become all the same. 
This proves the claim.

Second, we can derive an alternative formulation of Observation~1, 
as one can compute the geometric measure by finding the nearest 
separable  state. 
If we consider an entangled state with $f^+ > f^-$ it follows that the 
closest separable state is at the border of the deltoid, where 
$f^+ = 1/4 + f^-/2.$ Since the corresponding density matrices $\vr$
and $\sigma$ are diagonal in the same basis, the Uhlmann fidelity 
can easily be evaluated and minimized over this line. This leads to:

{\bf Observation 4.}
For entangled GHZ-symmetric states of three qubits with $f^+ > f^-$ 
the geometric measure is given by
\begin{align}
E_G^{(3)}(\vr) = &1 - \max_{\mu \in [0,1/2]}
\Big[
\frac{1}{4}(\sqrt{3}\sqrt{1-f^+-f^-}\sqrt{1-2\mu}
\nonumber \\
&+
2\sqrt{f^-}\sqrt{\mu}+\sqrt{f^+}\sqrt{1+2\mu})^2
\Big].
\label{finalopti-o4}
\end{align}
Again, this formula can in principle be evaluated for any 
values of $f^+, f^-$, but the solution then becomes technical. 
We stress, however, a crucial difference between Eq.~(\ref{finalopti-o1})
and Eq.~(\ref{finalopti-o4}): Eq.~(\ref{finalopti-o1})
holds for all values $f^+ > f^-$, consequently the value vanishes 
for separable states. Eq.~(\ref{finalopti-o4}) holds for states 
which are known to be entangled, but it is non-vanishing and not 
correct for separable states. A more practical difference is that
Eq.~(\ref{finalopti-o1}) seems to be better suited for obtaining
closed expressions as Eqs.~(\ref{upper-edge}, \ref{lowerborder})
than Eq.~(\ref{finalopti-o4}).

\section{Conclusions}
In conclusion, we have demonstrated how the geometric measure of 
entanglement and other entanglement measures can be evaluated for 
various  situations. This solves the problem of analyzing multiparticle
entanglement for GHZ-symmetric states. Our methods were 
enabled by connecting the usage of the Legendre transformation with
facts on the relation between entanglement measures and the Uhlmann 
fidelity. Our results will be useful in analyzing experiments aiming 
at the preparation of GHZ state, but also for analyzing the structure
of the different forms of multiparticle entanglement. A natural 
extension of our work would be the complete evaluation of an entanglement 
monotone for general states of three qubits, similar as it has been done 
for the entanglement of formation for two qubits \cite{wootters}. We hope that 
the presented results can be helpful for this task. 

We thank Aditi Sen(De) and Ujjwal Sen for discussions.
This work has been supported by  
the BMBF (Chist-Era Project QUASAR),
the DFG, 
the EU (Marie Curie CIG 293993/ENFOQI), 
the FQXi Fund (Silicon Valley Community Foundation),
and J.S.~Bach (BWV 248).

\section{Appendix}
\subsection{Proof of Observation 1}

From the discussion in the main text it is clear that
computing the geometric measure is equivalent to 
computing the optimal lower bound on the geometric measure 
for an arbitrary quantum state of which only the fidelities 
$f^+$ and $f^-$ are known. To do so, we use the method of the 
Legendre transformation. In Ref.~\cite{legpra} it has been shown 
that we have first to compute the Legendre transform
of the geometric measure, 
\be
\hat E (\mu \WW_+, \nu \WW_-) = \sup_\vr \{ Tr[\vr (\mu \WW_+ + \nu \WW_-)]
-E_G^{(3)}(\vr)\},
\label{legendre}
\ee
where $\WW_\pm = \ketbra{GHZ^\pm}$ are the operators measuring the 
fidelities $f^+$ and $f^-$. Given the Legendre transform the 
optimal lower bound based on the fidelities is then given by 
\be
E_G^{(3)}(f^+, f^-) = \sup_{\mu, \nu} \{ \mu f^+ + \nu f^- - 
\hat E (\mu \WW_+, \nu \WW_-)\}.
\ee
For the geometric measure, the computation of the Legendre transform
can be simplified \cite{legpra}: First, in Eq.~(\ref{legendre}) it suffices 
to optimize over pure states only since the measure is defined via 
the convex roof. Then, we can insert the definition of the geometric 
measure as a supremum over all pure product states. So we have to solve
\be
\hat E (\mu,\nu) = \sup_{\ket{\psi}} \sup_{\ket{\phi} \in \rm pr.~stat.}
\{ \bra{\psi} X \ket{\psi} + |\braket{\phi}{\psi}|^2-1,
\}
\ee
where $\ket{\phi}$ denotes a product state, 
and $X=\mu \ketbra{GHZ^+} +\nu \ketbra{GHZ^-}.$ 

To proceed, we expand everything in the eight basis vectors
$\ket{GHZ_i}$ of the GHZ basis \cite{duer01, kay}, where we set  
$\ket{GHZ^\pm}= \ket{GHZ_{1/2}}$. We write 
$\ket{\psi}= \sum_i \beta_i \ket{GHZ_i}$ and 
$\ket{\phi}= \sum_i \alpha_i \ket{GHZ_i}$, and 
$X= \sum_i \lambda_i \ketbra{GHZ_i}$ 
is also diagonal in this basis. We need to 
maximize
\be
\FF = |\sum_i \alpha_i \beta_i |^2+\sum \lambda_i|\beta_i|^2.
\ee
When solving this optimization problem, we have to take care of the 
conditions on the $\alpha_i$ and $\beta_i$. Especially the $\alpha_i$ 
obey several 
constraints besides the trivial normalization, as $\ket{\phi}$ is a 
product vector. In the following, we will relax the constraints at some 
point but in the end the solution of the relaxed problem will also be a 
solution of the original problem.

Since we want to maximize $\FF$, we can without loosing generality assume that all 
the coefficients are real and positive, and write 
$\FF = (\sum_i \alpha_i \beta_i)^2+\sum \lambda_i \beta_i^2$.
Then we start by characterizing the optimum of this function in some 
more detail. When $\FF$ is maximal, there are several possible 
transformations of the $\alpha_i$ and $\beta_i$, which can not 
increase the value of $\FF$ and this characterizes the optimum.

\begin{figure}[t!]
\begin{center}
\includegraphics[width=0.9\columnwidth]{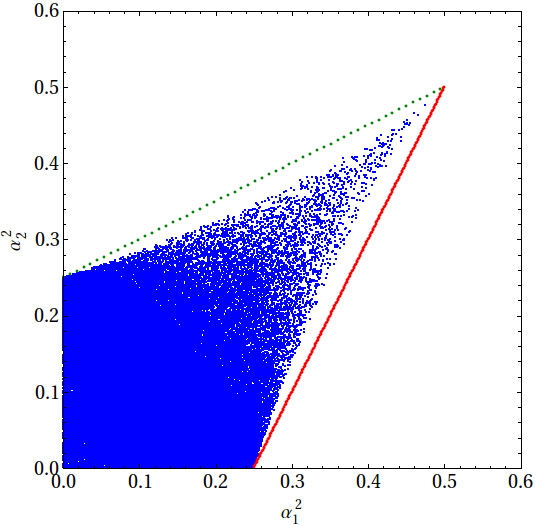}
\end{center}
\caption{The possible values of $\alpha_1^2$ and $\alpha_2^2$. The 
blue points are values from randomly generated pure product states
and the red solid and green dotted lines are the conditions from 
Eq.~(\ref{alphacondition}).
}
\label{eg-bild-1}
\end{figure}

A {\it first possible transformation} is the following: One may 
choose
coefficients $x_i$ with $\sum_i x_i =0$ and a small $\varepsilon$ 
and consider the map $\beta_i \mapsto \sqrt{\beta_i^2+\varepsilon x_i}.$
This keeps the normalization $\sum_i \beta_i^2=1$ and it should not
increase the value at the maximum. Taking the derivative with respect to
$\varepsilon$ and setting afterwards $\varepsilon=0$ 
we find that at the maximum
\be
(\sum_i \alpha_i \beta_i) \sum_k x_k \frac{\alpha_k}{\beta_k} 
+ \sum_l \lambda_l x_l = 0 
\label{firsttrafo}
\ee
holds, for any choice of the $x_i$. In other words, we can state that at the optimum
$\sum_i \lambda_i x_i > 0$ is equivalent to $\sum_i x_i (\alpha_i/\beta_i) <0.$

A {\it second possible transformation} affects the $\alpha_i$. For the map
$\alpha_i \mapsto \sqrt{\alpha_i^2+\varepsilon x_i}$ the same argument
as above gives that 
\be
(\sum_i \alpha_i \beta_i) \sum_k x_k \frac{\beta_k}{\alpha_k} = 0
\label{secondtrafo}
\ee
holds at the optimum. Since for the optimal solution there is a finite overlap
between $\ket{\psi}$ and $\ket{\phi}$ it follows that
$\sum_k x_k {\beta_k}/{\alpha_k} = 0$.
Note that this second transformation might be in conflict with the 
conditions on the $\alpha_i$ coming from the product structure of the 
vector $\ket{\phi}.$ However, if the coefficients $\alpha_i$ are not at 
the boundary of the allowed values, we can apply it.

As a second step, we have to characterize the possible values of 
$\alpha_i$ in some more detail. More precisely, we claim and will 
use that (see also Fig.~\ref{eg-bild-1})
\be
\alpha_1^2 \leq \frac{1}{4}+\frac{\alpha_2^2}{2}
\quad\mbox{  and  }\quad
\alpha_2^2 \leq \frac{1}{4}+\frac{\alpha_1^2}{2}.
\label{alphacondition}
\ee
This follows from the product structure of $\ket{\phi}$ in the following
way: It has been shown that the density matrix elements $\vr_{i,j}$ of any 
fully separable three-qubit state obey \cite{seevinck}
\be
|\vr_{1,8}| \leq \sqrt[6]{\vr_{2,2}\vr_{3,3}\vr_{4,4}\vr_{5,5}\vr_{6,6}
\vr_{7,7}},
\label{sepcond}
\ee
that is, the off-diagonal element $|\vr_{18}|$ is bounded by a 
function of the diagonal elements. For the product vector $\ket{\phi}$
we have that $|\vr_{1,8}|=|\alpha_1^2-\alpha_2^2|/2$ and 
$\vr_{2,2}=(\alpha_3^2+\alpha_4^2)/2$ etc., and maximizing  the 
right hand side of Eq.~(\ref{sepcond}) under the normalization 
$\sum_i \alpha_i^2 = 1$, one can directly see that 
Eq.~(\ref{alphacondition}) holds. We note that there are further 
constraints on $\alpha_1^2$ and $\alpha_2^2$ from the product structure, 
but the conditions above together with the positivity of the $\alpha_1^2$ 
and $\alpha_2^2$ represent already the convex hull of all 
possible values, see also Fig.~\ref{eg-bild-1}.

After this preliminary considerations we are ready to tackle 
the main optimization problem of finding the maximum of
$\FF = (\sum_i \alpha_i \beta_i)^2+\sum_i \lambda_i \beta_i^2$. 
We first assume that $\lambda_1 =\mu > 0$ and $\lambda_2 = \nu < 0$, 
the other $\lambda_i$ vanish in any case. All the contributions in 
the zero space of $X$ can be subsumed by single numbers $\alpha_3$ and 
$\beta_3$, so without loosing generality we consider only three indices 
$i=1,2,3.$ We also write the coefficients as vectors 
$\vec{\alpha}=(\alpha_1, \alpha_2, \alpha_3)$ etc., and denote their 
scalar product as $\braket{\vec{\alpha}}{\vec{\beta}}.$ Our aim is to 
show that the optimal solution of the product vector $\ket{\phi}$ is 
either given by $\alpha_1^2=1/4$ and $\alpha_2^2=0$ (corresponding to 
the product vector $\ket{\phi}=\ket{+}\ket{+}\ket{+}$ with 
$\ket{+}=(\ket{0}+\ket{1})/\sqrt{2}$) or by 
$\alpha_1^2=\alpha_2^2=1/2$ (corresponding to the product vector 
$\ket{\phi}=\ket{000}$).

As the $\beta_i$ have to fulfill only the normalization constraint 
and $\mu >0$ it is clear that for a given value of $\alpha_2^2$ the choice of a
maximal $\alpha_1^2$ is optimal. This means that we only have to 
consider the line $\alpha_2^2 = 2 \alpha_1^2- 1/2$  with
$\alpha_1^2 \in [1/4;1/2]$, being the red solid line in Fig.~\ref{eg-bild-1}. 
The endpoints of this line are the possible solutions 
mentioned above. Note that at this line we have 
$\alpha_2 =\sqrt{2 \alpha_1^2 - 1/2}$ and due to the 
normalization $\alpha_3=\sqrt{3/2-3\alpha_1^2}$, so everything can 
be expressed in terms of $\alpha_1.$

Let us assume that the solution is not at the endpoints. In the following 
we will show that under this condition the vector $\vec{\beta}$ is already 
determined, and the optimal value can be computed. It turns out, 
however, that the optimum is then the same for all values $\vec{\alpha}$
on the line, so we can also choose the endpoints.

First we can take the derivative in direction of this line. This is done by choosing
$\vec{x}=(x_1,x_2, x_3) = (1, 2, -3)$ and applying the second transformation
from above. It follows that 
\be
\frac{\beta_1}{\alpha_1}+ \frac{2\beta_2}{\alpha_2}- \frac{3\beta_3}{\alpha_3}
=0.
\label{betacond1}
\ee
So, if we define the vector
\be
\vec{v}= \NN (\frac{1}{\alpha_1},\frac{2}{\alpha_2},\frac{-3}{\alpha_3}),
\ee
where $\NN$ denotes a normalization, we have that 
$\braket{\vec{\beta}}{\vec{v}}=0$ and $\braket{\vec{\alpha}}{\vec{v}}=0.$

Then we consider a transformation of the first type from above, 
affecting the $\beta_i$. For a general vector $\vec{x}$ we have 
that Eq.~(\ref{firsttrafo}) holds. Especially, we may choose 
$x_i=v_i \beta_i$, then also $\sum_i x_i =0$ is satisfied. Then
we have for the first term in Eq.~(\ref{firsttrafo}) 
$\sum_k x_k \alpha_k/ \beta_k = \sum_k v_k \alpha_k =0$ and 
it follows that 
\be
\beta_1 \frac{\lambda_1}{\alpha_1}+2\beta_2 \frac{\lambda_2}{\alpha_2}=0.
\label{betacond2}
\ee
Equations (\ref{betacond1}, \ref{betacond2}) allow to determine $\vec{\beta}$
up to the normalization. Choosing $\beta_1=1$ we find that
\be
\beta_2=-\frac{\sqrt{2\alpha_1^2-1/2} \lambda_1}{2\alpha_1 \lambda_2},
\;\;\;
\beta_3=\frac{\sqrt{1-2\alpha_1^2}(\lambda_2-\lambda_1)}{\sqrt{6} \alpha_1 \lambda_2}.
\ee
We can insert this solution again in Eq.~(\ref{firsttrafo}) and use
that this holds still for all admissible $\vec{x}=(x_1, x_2, -x_1-x_2).$
This leads to three possible solutions of $\lambda_2$ as a function of 
$\lambda_1$. From the conditions $\lambda_1 >0 >\lambda_2$ 
it then follows that
\be
\lambda_2= \frac{\lambda_1^2+\lambda_1\sqrt{(1-\lambda_1+\lambda_1^2)}}{2(\lambda_1-1)}
\ee
has to hold, with $\lambda_1 \in [0,1).$ For these values of
$\lambda_1, \lambda_2,$ however, one directly finds that the 
function $\FF$ does not depend on $\alpha_1^2$, so $\FF$ is 
constant on the red solid line in Fig.~\ref{eg-bild-1}. In this way, 
we have shown that the only relevant points are $\alpha_1^2=1/4$ and 
$\alpha_1^2=1/2.$ 

Having fixed $\vec{\alpha}$, the optimal $\vec{\beta}$ can directly
be determined by looking for the maximal eigenvalue of a $3\times3$
matrix. In this way one determines the Legendre transform as
\begin{align}
\hat{E}(\mu, \nu) &=\frac{1}{2} \big[\mu-1+\sqrt{1+\mu(\mu-1)^2}\big]
\label{ehat1}
\\
& \mbox{for } \mu \in (0,1) 
\mbox{ and }\nu <  \frac{\mu^2+\mu\sqrt{(1-\mu+\mu^2)}}{2(\mu-1)},
\nonumber
\\
\hat{E}(\mu, \nu) &=\frac{1}{2} \big[\mu +\nu -1+\sqrt{1-(\mu-\nu)^2}\big]
\label{ehat2}
\\
& \mbox{for all other values } \mu > 0 \mbox{ and } \nu < 0 .
\nonumber
\end{align}

This solves the first problem for $\mu>0$ and $\nu<0$, let us shortly 
discuss what happens for the other possible signs: $\mu<0$ and $\nu>0$
can directly be solved as above. For the case $\mu >0, \nu >0$ the solution
is obviously given by choosing $\alpha_1^2=\alpha_2^2=1/2$, leading to 
the formula (\ref{ehat2}) again. Finally, if $\mu <0, \nu < 0$ the optimum is
attained by choosing  $\alpha_1^2=\alpha_2^2=0$ and $\alpha_3^2=1$. But then, 
$\hat{E}(\mu, \nu)=0$ does not depend on $\mu$ and $\nu$ at all,
so this case has no physical relevance.

It remains to compute the geometric measure from the Legendre transform
and the values of the fidelity. We have to optimize
\be
E(f^+, f^-) = 
\sup_{\mu, \nu} \{ \mu f^+ + \nu f^- - \hat E (\mu, \nu)\}
\ee
for the functions given in Eqs.~(\ref{ehat1}, \ref{ehat2}). Let us 
consider the parameters as in Eq.~(\ref{ehat1}) first. In this case, 
$\hat E (\mu, \nu)$ does not depend on $\nu$ at all, so it is clear
to take $\nu$ as large as possible, since $f^-\geq 0$. So we have to 
choose $\nu = [{\mu^2+\mu\sqrt{(1-\mu+\mu^2)}}]/[{2(\mu-1)}].$ 
If $\hat{E}$ is given by Eq.~(\ref{ehat2}), we can consider a 
transformation $\mu \mapsto \mu - \varepsilon$ and 
$\nu \mapsto \nu - \varepsilon$. This increases $(-\hat{E})$ by 
$\varepsilon$, on the other hand, the term $\mu f^+ + \nu f^-$ 
decreases less, since $f^+ + f^- \leq 1.$ Consequently, it is optimal 
to take $\varepsilon$ as large as possible, which means  that
the optimum is attained again at the boundary, where
$\nu = [{\mu^2+\mu\sqrt{(1-\mu+\mu^2)}}]/[{2(\mu-1)}].$
Note that at this boundary the condition $\mu \in (0,1)$
is automatically fulfilled.

In summary, the computation of $E(f^+, f^-)$ can be reduced to an 
optimization over the single parameter $\mu,$ namely
\be
E_G^{(3)}(f^+, f^-) = 
\!\!
\max_{\mu \in [0,1]}
\frac{1}{2}
\big[
1+\mu (2 f^+ - 1)
-\sqrt{\alpha}+\frac{f^- \mu (\mu+\sqrt{\alpha})}{\mu-1}
\big],
\label{finalopti}
\ee
where we used the abbreviation $\alpha=1-\mu+\mu^2.$ This is the statement
of Observation 1. Clearly, the formula (\ref{finalopti}) gives the value 
$E_G^{(3)}$ for the case $f^+ \geq f^-$ only, since for this case 
$\mu>0>\nu$ is the relevant choice of  parameters. For the other case 
one has to exchange the values of $f^+$ and $f^-$.
\qed

\subsection{The optimal decomposition}

Here we find the optimal decomposition for GHZ-symmetric 
three-qubit states at the lower cathetus where $f^-=0.$  
Let us denote these states by $\vr(f^+)$. The entries of these  
density matrices are given by
$\vr_{1,1}=\vr_{8,8}=\vr_{1,8}=\vr_{8,1}=f^+/2$, 
the other entries on the diagonal are 
$\vr_{2,2}=\vr_{3,3}=...=\vr_{7,7}=(1-f^+)/6$ and all
other entries vanish. For the decomposition, we consider first 
the following four vectors:
\begin{align}
\ket{\psi_1} &= \alpha (\ket{000}+\ket{111})
\nonumber
\\
&+ \beta (\ket{001}+\ket{010}+\ket{011}+\ket{100}+\ket{101}+\ket{110}),
\nonumber
\\
\ket{\psi_2}&=\sigma_z \otimes \sigma_z \otimes \eins \ket{\psi_1},
\nonumber
\\
\ket{\psi_3}&=\sigma_z \otimes \eins \otimes \sigma_z  \ket{\psi_1}, 
\nonumber
\\
\ket{\psi_4}&=\eins \otimes \sigma_z \otimes \sigma_z \ket{\psi_1},
\label{firstset}
\end{align}
with $\alpha = \sqrt{f^+/2}$ and $\beta=\sqrt{(1-2\alpha^2)/6}
=\sqrt{(1-f^+)/6}.$ The density matrix $\ketbra{\psi_1}$ has the same
entries on the diagonal as $\vr(f^+)$ and the entries 
$\vr_{1,8}$ and $\vr_{8,1}$ also coincide. The symmetrized density 
matrix $X_1=(1/4)\sum_{k=1}^4 \ketbra{\psi_k}$ fulfills the same properties, 
but has only elements on the diagonal and anti-diagonal. 

So far, the values of $X_1$ on the anti-diagonal other than
$\vr_{1,8}$ and $\vr_{8,1}$ do not vanish, contrary to the 
values of $\vr(f^+)$. So we consider vectors of the type
\be
\ket{\xi(\phi_1,\phi_2,\phi_3)}=
e^{i \phi_1 \sigma_z} \otimes e^{i\phi_2 \sigma_z} \otimes e^{i\phi_3\sigma_z}
\ket{\psi_1}
\ee
with the constraint $\phi_1 + \phi_2 + \phi_3 = 0.$ As explained before, 
this is also a symmetry of the states in the considered family.
We first consider 
\be
\ket{\psi_5}= \ket{\xi(\pi/4,\pi/4,-\pi/2)}
\ee
and define $\ket{\psi_6}, \ket{\psi_7}, \ket{\psi_8}$ via a 
symmetrization as in Eq.~(\ref{firstset}). The matrix
$X_5=(1/4)\sum_{k=5}^8 \ketbra{\psi_k}$ is similar to
$X_1$, but the entry $\vr_{2,8}$ has a different sign, and
$\vr_{3,7}$ and $\vr_{4,5}$ have the phase $-i$. Adding the 
vectors
\be
\ket{\psi_9}= \ket{\xi(-\pi/4,-\pi/4,\pi/2)}
\ee
with the symmetrizations $\ket{\psi_{10}}, \ket{\psi_{11}}, \ket{\psi_{12}}$
and $X_9=(1/4)\sum_{k=9}^{12} \ketbra{\psi_k}$ we find that
$Y = X_1/2 + X_5/4 + X_9/4$ has, apart from $\vr_{3,7}$ and $\vr_{4,5}$ 
all the entries as $\vr(f^+)$, also $\vr_{2,8}$ is vanishing.
Repeating this procedure with
\bea
\ket{\psi_{13}} &=& \ket{\xi(\pi/4,-\pi/2,\pi/4,)},
\nonumber
\\
\ket{\psi_{17}} &=& \ket{\xi(-\pi/4,\pi/2,-\pi/4)},
\nonumber
\\
\ket{\psi_{21}} &=& \ket{\xi(-\pi/2,\pi/4,\pi/4)},
\nonumber
\\
\ket{\psi_{25}} &=& \ket{\xi(\pi/2, -\pi/4,-\pi/4,)}
\eea
and the respective symmetrizations we arrive at a decomposition
of $\vr(f^+)$ into 28 pure states.

It remains to show that this is the optimal decomposition.
By taking the overlap with the product state 
$\ket{\Phi}=\ket{+}\ket{+}\ket{+}$ the geometric measure 
of $\ket{\psi_1}$ is bounded by 
\bea
E(\ket{\psi_1}) &\leq& 1- \frac{1}{8}(2\alpha+6\beta)^2
\nonumber
\\
&=&
\frac{1}{4}(1+2f^+-2\sqrt{3}\sqrt{f^+(1-f^+)})
\eea
which coincides with formula Eq.~(\ref{lowerborder}) if the fidelity
$f^+ \leq 3/4$. This also proves that in this regime the state 
$\ket{\Phi}$ was indeed the closest product state to $\ket{\psi_1}.$
For fidelities larger than that one can directly check that $\ket{\Phi}$
is not the closest state anymore, computing the geometric measure leads
then to a function that is not convex in $f^+$. Since the geometric measure
must be convex, however, the optimal decomposition is then not given
by the decomposition into the $\ket{\psi_1}$ anymore. Instead, the optimal
decomposition is for $f^+ > 3/4$ just given by
\be
\vr(f^+)=(1-p) \vr(f^+=3/4) + p \ketbra{GHZ^+}
\ee
with $p = 4f^+-3.$ Of course, $\vr(f^+=3/4)$ should be understood here as a
placeholder of its optimal decomposition into 28 pure states. 
This reproduces the second (linear) part of Eq.~(\ref{lowerborder}).
\qed

\subsection{Proof of Observation 2}
In order to show this Observation let us first
consider full separability of an $N$-qubit state (that is, $k=N$)
and discuss what parts of the previous proof of Observation 1 
require modification. The first modification is Eq.~(\ref{sepcond})
which has to be replaced by the generalization
\be
|\vr_{1, 2^N}| \leq (\vr_{2,2}\vr_{3,3}\cdot ... \cdot \vr_{2^N-1,2^N-1})^{\frac{1}{2^N-2}},
\label{sepcond2}
\ee
The corresponding generalization of Eq.~(\ref{alphacondition}) reads
\bea
\alpha_1^2 \leq \frac{1}{2^{N-1}}\big[1+(2^{N-1}-2){\alpha_2^2}\big],
\nonumber
\\
\alpha_2^2 \leq \frac{1}{2^{N-1}}\big[1+(2^{N-1}-2){\alpha_1^2}\big].
\label{alphacondition2}
\eea
In this way, the deltoid in Fig.~\ref{eg-bild-1} becomes smaller. This time, 
the state $\ket{\phi}=\ket{+}^{\otimes N}$ corresponds to the corner
$\alpha_1^2=1/(2^{N-2})$ and $\alpha_2^2=0$
and the state $\ket{\phi}=\ket{0}^{\otimes N}$ corresponds to 
$\alpha_1^2=\alpha_2^2=1/2.$

With this modification, one can first show as before that only the corners 
of the modified deltoid are relevant. Then, one has to diagonalize again $3\times3$ 
matrices to arrive at similar expressions to Eqs.~(\ref{ehat1}, \ref{ehat2}).
In fact, Eq.~(\ref{ehat2}) does not change, while Eq.~(\ref{ehat1}) reads
\begin{align}
\hat{E}(\mu, \nu) &=\frac{1}{2} \big[\mu-1+\sqrt{(\mu-1)^2+2^{3-N}\mu}\big].
\label{ehat1neu}
\end{align}
With this one can argue as before that only the line where the two types 
of the Legendre transform coincide is relevant. This leads to Eq.~(\ref{finaloptimulti}).

Finally, we have to discuss what happens if we consider the geometric measure for 
$k$ separability of an $N$ qubit state, if $k < N.$ This can be understood as 
follows: Let us assume that we have a pure $k$-separable state, which has to 
be separable for a fixed partition. To be explicit, consider the three-separable 
six-qubit state  
$\ket{\Psi}=\ket{\phi_1} \otimes \ket{\phi_{23}} \otimes \ket{\phi_{456}}$. 
We claim that this state obeys the conditions of Eq.~(\ref{alphacondition2})
with $N$ set to $k$. Indeed, by projecting the possibly entangled groups of 
$\ell$ qubits on the space $\Pi_\ell=\ketbra{00...0} + \ketbra{11...1}$ and 
identifying the logical qubits $\ket{0}_L=\ket{00...0}$ and $\ket{1}_L=\ket{11...1}$ 
we arrive at a state which is effectively a fully separable $k$-qubit state. In 
our example, we have to apply $\eins \otimes \Pi_2 \otimes \Pi_3$ to $\ket{\Psi}$ 
to arrive at this state. The fidelities of the GHZ states $\ket{GHZ^\pm}$ do not 
change, as the GHZ states are invariant under the projections. It follows that 
any pure $k$-separable  $N$-qubit state obeys the conditions of 
Eq.~(\ref{alphacondition2}) with $N$ set to $k$. In addition, one can also reach 
the corners of the deltoid by appropriate $N$-qubit $k$-separable states. From 
this point, the proof can proceed as before and one arrives at 
Eq.~(\ref{finaloptimulti}).
\qed

\subsection{Proof of Observation 3}
In this case, the deltoid from Fig.~\ref{eg-bild-1} becomes a square. 
When computing the eigenvalues of the matrices before Eqs.~(\ref{ehat1}, \ref{ehat2}) 
it is then clear that one of the possible solutions is larger than the other. So 
the Legendre transform reduces to 
\begin{align}
\hat{E}(\mu, \nu) &=\frac{1}{2} \big[\mu-1+\sqrt{1+\mu^2}\big].
\label{ehat1bi}
\end{align}
In the final optimization, one has to set $\nu=0.$ The remaining optimization
over $\mu \geq 0$ can be carried out analytically, and one finds that if 
$f^+\geq 1/2$ the geometric measure is given by
\be
E(f^+, f^-) = \frac{1}{2}-\sqrt{f^+(f^+-1)}.
\ee
The analogous formula holds if $f^-\geq 1/2$. For the other cases
the geometric measure vanishes, as the state is biseparable.
\qed


\begin{thebibliography}{99}

\bibitem{gtreview}
R.~Horodecki, P.~Horodecki, M.~Horodecki, and K.~Horodecki,
{Rev. Mod. Phys. \textbf{81}, 865 (2009)},
O.~G\"uhne and G.~T\'oth,
{Phys. Rep. \textbf{474}, 1 (2009)}.

\bibitem{measure-reviews}
M. B. Plenio and S. Virmani,
Quantum Inf. Comput. {\bf 7}, 1 (2007);
C. Eltschka and J. Siewert,
J. Phys. A: Math. Theor. {\bf 47}, 424005 (2014). 

\bibitem{wei}
T.-C.~Wei and P.M.~Goldbart,
Phys.~Rev.~A {\bf 68}, 042307 (2003), 
see also A. Shimony, Ann. N.Y. Acad. Sci. {\bf 755}, 
675 (1995).

\bibitem{illuminati}
M. Blasone, F. Dell’Anno, S. De Siena, and F. Illuminati
Phys. Rev. A {\bf 77}, 062304 (2008);
A. Sen De and  U. Sen,
Phys. Rev. A {\bf 81}, 012308 (2010).

\bibitem{geo-apps}
M.~Hayashi \emph{et al.}, D.~Markham, M.~Murao, M.~Owari, and S.~Virmani,
Phys.~Rev.~Lett.~{\bf 96}, 040501 (2006).


\bibitem{geo-spin}
T.-C.~Wei, 
D.~Das, S.~Mukhopadyay, S.~Vishveshwara, and P.M.~Goldbart,
Phys.~Rev.~A {\bf 71}, 060305(R) (2005);
R.~Orus,
Phys.~Rev.~Lett.~{\bf 100}, 130502 (2008);
H. Shekhar Dhar, A. Sen De, and  U. Sen,
Phys. Rev. Lett. {\bf 111}, 070501 (2013).

\bibitem{streltsov}
A. Streltsov, H. Kampermann, and D. Bru{\ss},
New J. Phys. {\bf 12}, 123004 (2010).

\bibitem{gbb}
O. G\"uhne, F. Bodoky, and M. Blaauboer,
Phys. Rev. A {\bf 78}, 060301 (2008).

\bibitem{indiaposter}
S. Bagchi, A. Misra, A. Sen(De), and U. Sen,
poster at the IWQI 2012 workshop in Allahabad, 
India (2012).

\bibitem{wootters}
W. K. Wootters,
{Phys. Rev. Lett.} {\bf 80}, 2245 (1998);
A. Uhlmann,
Phys. Rev. A {\bf 62} 032307 (2000);
Entropy {\bf 12}, 1799 (2010).

\bibitem{eltschka}
C. Eltschka and J. Siewert,
Phys. Rev. Lett. {\bf 108}, 020502 (2012).

\bibitem{furtherexact}
B. M. Terhal and K. G. H. Vollbrecht,
{Phys. Rev. Lett.} {\bf 85}, 2625 (2000);
K. G. H. Vollbrecht and R. F. Werner,
{Phys. Rev. A} {\bf 64}, 062307 (2001);
P. Rungta and C. M. Caves,
{Phys. Rev. A} {\bf 67}, 012307 (2003);
R. Lohmayer, A. Osterloh, J. Siewert, and A. Uhlmann,
Phys. Rev. Lett. {\bf 97}, 260502 (2006).

\bibitem{geza-neu}
G T\'oth, T. Moroder, and O. G\"uhne,
arXiv:1409.3806.

\bibitem{horosym}
K. Horodecki, M. Horodecki, and P. Horodecki, 
Quantum Inf. Comput. {\bf 10}, 901 (2010). 

\bibitem{eltschka2}
C. Eltschka and J. Siewert, 
Phys. Rev. A {\bf 89}, 022312 (2014).

\bibitem{legendretrafo}
O. G\"uhne, M. Reimpell, and R.F. Werner, 
Phys. Rev. Lett. {\bf 98}, 110502 (2007);
J. Eisert, F. Brand\~ao, and K. Audenaert, 
New J. Phys. {\bf 9}, 46 (2007).

\bibitem{compi} We {strongly} recommend the use of a 
computer algebra package at this point.

\bibitem{duer01}
W. D{\"u}r and J. I. Cirac,
J. Phys. A: Math. Gen. {\bf 34}, 6837 (2001). 

\bibitem{kay}
A. Kay, Phys. Rev. A {\bf 83}, 020303(R) (2011).

\bibitem{martin}
J. Martin, O. Giraud, P. A. Braun, D. Braun, and T. Bastin,
Phys. Rev. A {\bf 81}, 062347 (2010).

\bibitem{karol}
R. Jozsa, J. Mod. Optics {\bf 41}, 2315 (1994).

\bibitem{legpra}
O. G\"uhne, M. Reimpell, and R.F. Werner,
Phys. Rev. A. {\bf 77}, 052317 (2008).

\bibitem{seevinck}
O. G\"uhne and M. Seevinck, 
New J. Phys. {\bf 12}, 053002 (2010).


\end{thebibliography}
\end{document}